\documentclass[twocolumn,floatfix,showpacs,pre]{revtex4}

\usepackage{epsfig}
\usepackage{amsmath}
\usepackage{times}
\usepackage{color}

\begin{document}

\title{Spectral representation of the effective dielectric constant of  graded composites}

\author{L. Dong}
\affiliation{Biophysics and Statistical Mechanics Group, 
Laboratory of Computational Engineering,
Helsinki University of Technology, P.\,O.~Box 9203, FIN-02015 HUT, Finland} 
\affiliation{Department of Physics, The Chinese University of Hong Kong, 
Shatin, New Territories, Hong Kong}
\author{Mikko Karttunen}
\affiliation{Biophysics and Statistical Mechanics Group, 
Laboratory of Computational Engineering,
Helsinki University of Technology, P.\,O.~Box 9203, FIN-02015 HUT, Finland}
\author{K. W. Yu}
\affiliation{Department of Physics, The Chinese University of Hong Kong, 
Shatin, New Territories, Hong Kong}

\date{\today}

\begin{abstract}

We generalize the Bergman-Milton spectral representation, originally
derived for a two-component composite, to extract the spectral density
function for the effective dielectric constant of a graded composite.  
This work has been motivated by a recent study of the optical absorption
spectrum of a graded metallic film (Applied Physics Letters, {\bf 85}, 94
(2004)) in which a broad surface-plasmon absorption band has been shown
to be responsible for enhanced nonlinear optical response as well as an
attractive figure of merit.  It turns out that, unlike in the case of
homogeneous constituent components, the characteristic function of a
graded composite is a continuous function because of the continuous
variation of the dielectric function within the constituent components.
Analytic generalization to three dimensional graded composites is
discussed, and numerical calculations of multilayer composites are given
as a simple application.

\end{abstract}

\maketitle

\section{Introduction}

In graded materials~\cite{Milton1} the physical 
properties may vary continuously in space making them distinctly 
different from homogeneous materials. Hence,
composite media consisting of graded inclusions 
have attracted much interest in various engineering
applications~\cite{Yamanouchi}, such as reduced residual and thermal
barrier coatings of high temperature components in gas turbines, 
surface hardening for tribological protection, and graded interlayers used
in multilayered microelectronic and optoelectronic
components~\cite{Holt,Cherradi}.

Like graded materials, thin films are of great interest in many practical applications
and often possess different optical
properties~\cite{Kammler} in comparison to bulk materials.
Recently, it was found experimentally that graded thin films may have
high relative dielectric permittivity as well as a flatter temperature
characteristic of permittivity~\cite{Kawasaki} than single-layer
films~\cite{LuAPL03}.  

The traditional theories used to deal
with the homogeneous materials~\cite{Jackson,AIP}, however fail to deal
with composites of graded inclusions. To treat these composites, we
have recently developed a first-principles approach~\cite{Dong,Gu-JAP}
and a differential effective dipole approximation~\cite{Huang}.

This work has been motivated by a recent study of the optical absorption
spectrum of a graded metallic film~\cite{Huang1}. In that work, a broad
surface plasmon absorption band was observed in addition to a strong
Drude absorption peak at zero frequency. Such a broad absorption band has
been shown to be responsible for the enhanced nonlinear optical response
as well as an attractive figure of merit (the degree of optical
absorption). Yuen {\em et al.}~\cite{Yuen1} pointed out that such an
absorption spectrum, being related to the imaginary part of the effective
dielectric constant, should equally well be reflected in the
Bergman-Milton spectral representation of the effective dielectric
constant~\cite{Bergman,Milton}.

Bergman-Milton spectral representation was originally developed for calculating the
effective dielectric constant and other response functions of
two-component composites~\cite{Bergman,Milton}. However, the two
concerned components are all homogeneous. Therefore, it is worth
extending the spectral representation to graded composite materials. The
work on graded films is just a simple example of a more general graded
composite in three dimensions. One of the main purpose of this work help
to identify the physical origin of the broad absorption band. It turns
out that, unlike in the case of homogeneous materials, the characteristic
function of a graded composite is a continuous function because of the
continuous variation of the dielectric function within the constituent
component.

Moreover, we apply our theory to a special case of graded composites,
i.\,e., multilayer material, which is more convenient to fabricate in
practice than graded material~\cite{Hobson}, and many algorithms are now
available for designing of multilayer coatings~\cite{Martin,Verly}. Thus,
the present work is necessary in the sense that we shall discuss the
multilayer effect as the number of layers inside the material increases.
In this regard, this work should be expected to have practical relevance.
As the number of layers $N$ increases, we shall show a gradual transition
from sharp peaks to a broad continuous band until the graded composite
results are recovered by the limit of $N\to \infty$.

The paper is organized as follows. In Sec.~\ref{sec:forma}, the general derivation
of the spectral representation for graded composites is presented. In
Sec.~\ref{sec:spectral}, we describe the model and present analytical results for the
spectral representation of the effective dielectric constant of a graded
film with an interface, as well as a graded sphere. In those cases, the
Bergman-Milton formalism has been modified for graded composites. We
further obtain an analytic form for the spectral density function of a
multilayer film and a multilayer sphere in Sec.~\ref{sec:multi}. Numerical results
are presented in Sec.~\ref{sec:numerical}, and discussion and conclusion are given in
Sec.~\ref{sec:discussion}.

\section{Formalism \label{sec:forma}}

We consider a two-component composite in which graded inclusions of
dielectric constant $\epsilon_{1}({\bf r})$ are embedded in a homogeneous
host medium of dielectric constant $\epsilon_{2}$. It is noted that the
dielectric constant $\epsilon_{1}({\bf r})$ is a gradation profile as a
function of the position ${\bf r}$. And we will restrict our discussion
and calculation to the quasi-static approximation, i.\,e., $dc/{\omega}
\leq 1$, where $d$ is the characteristic size of the inclusion, $c$ is
the speed of light in vacuum and $\omega$ is the frequency of the applied
field. In the quasi-static approximation, the whole graded structure can
be regarded as an effective homogeneous one with effective (overall)
linear dielectric constant defined as~\cite{Stroud}
\begin{equation}
\epsilon_{e}=\frac{1}{V}\int \displaystyle \frac{{\bf E}\cdot {\bf D}}{E_{0}^{2}}dV,
\label{eq:1}
\end{equation}
where ${\mathit E}_{0}$ is the applied electric field along $z$
direction, ${\bf E}$ and ${\bf D}$ are the local electric field and local
displacement, respectively.

The object of the present section is to solve the Laplace's equation
\begin{equation}
{\bf \nabla}\cdot(\epsilon({\bf r})\nabla \phi({\bf r}))=0 
\label{eq:2}
\end{equation}
subject to the boundary condition $\phi_{0}=-{\mathit E}_{0}z$. The
dielectric function $\epsilon({\bf r})$ varies from component to
component but has a fixed mathematical expression for a given component.
It can be expressed as~\cite{Bergman}
\begin{equation}
\epsilon({\bf r})=\epsilon_{2}\left[1-\frac{1}{{\mathit s}}\eta({\bf r})\right],
\label{eq:3}
\end{equation}
where ${\mathit s}=[1-\epsilon_\mathrm{ref}/\epsilon_{2}]^{-1}$ is the
material parameter and $\epsilon_\mathrm{ref}$ is some reference
dielectric constant in the graded component. The characteristic function
$\eta({\bf r})$ is may be written in terms of a real function $f({\bf
r})$ as
\[
\eta({\bf r})=\left\{\begin{array}{cc}
1+f({\bf r}) \ \  & \text{ in inclusion}, \\
0 \ \ \ \ &  \text{in host},
\end{array} 
\right.
\]
which accords for the microstructure of graded composites. The function
$f({\bf r})$ depends on the specific variation of the dielectric constant
in the inclusion component. For homogeneous constituent component, i.\,e.,
$f({\bf r})=0$, $\eta({\bf r})=1$ in the inclusion component, while
$\eta({\bf r})=0$ in the host medium. For graded systems, $\eta({\bf r})$
can be a continuous function in the inclusion component because of the
continuous variation of the dielectric function within the inclusion
component.  Thus, Eq.~(\ref{eq:1}) can be solved
\begin{equation}\label{medbvp}
\phi({\bf r})=-{\mathit E}_{0}z+\frac{1}{s}\int 
d{\mathit V}^{\prime} \eta({\bf r}^{\prime})
{\nabla}^{\prime}G({\bf r}-{\bf r}^{\prime}) 
\cdot {\nabla}^{\prime} \phi({\bf r}^{\prime}),
\end{equation}
where $G({\bf r-r^\prime })$ is a Green's function satisfying:
\[
\left\{\begin{array}{cc}
  {\nabla}^2 G({\bf r- r^\prime })=-\delta^3({\bf r- r^\prime }) \ \ \text{for {\bf r} in
V},\\
  G=0 \ \text{ for {\bf r} on the boundary.}
 \end{array}
\right.
\]
In order to obtain a solution for Eq.~(\ref{eq:2}), we introduce an
integral-differential Hermitian operator $\hat {\Gamma}$, which satisfies
\[
\hat{\Gamma} \equiv \int d{\mathit V}^{\prime} 
\eta({\bf r}^{\prime}){\nabla}^{\prime}
G({\bf r}-{\bf r}^{\prime}) \cdot {\nabla}^{\prime},
\]
and define an inner product as 
\begin{equation}\label{scalPr}
 \left< \phi | \psi \right>= 
\int d {\mathit V} \eta({\bf r}) \nabla \phi^{*}\cdot \nabla \psi.
\end{equation}
With the above definitions, Eq.~(\ref{medbvp}) can be simplified to
\[
\phi({\bf r})= -E_{0} z +\frac{1}{s} \hat{\Gamma} \phi({\bf r}).
\]

Let $s_n$ and $|\phi_n \rangle$ be the $n$th eigenvalue and eigenfunction
of operator $ \hat{\Gamma}$. Then, the generalized eigenvalue problem
becomes
\[
{\bf \nabla}\cdot(\eta({\bf r})\nabla \phi_{n})={\mathit s}_{n} {\nabla}^{2}
\phi_{n}.
\]
The potential $|\phi \rangle$ can be expanded in series of eigenfunctions,
\begin{equation}\label{solution}
|\phi\rangle
       \equiv \sum_{n}  \left(\frac{s}{s_n-s}\right)
\frac{|\phi_n\rangle\langle\phi_n|z\rangle}{\langle\phi_n|\phi_n\rangle},
\end{equation}
where we choose $E_{0}=1$ for convenience.  Since $\eta({\bf r})$ is a
real function, the eigenvalues ${\mathit s}_{n}$ will be real. Also, for
graded component, $\eta({\bf r})$ is a continuous function, which will
cover the full region, i.\,e., $-\infty \le \eta({\bf r}) \le \infty$.
Therefore, the eigenvalues ${\mathit s}_{n}$, which depend on the
continuously graded microstructure $\eta({\bf r})$, do not lie within the
interval $[0,1]$ but extend to $-\infty \le s_n \le \infty$ as first
pointed by Gu and Gong~\cite{GuYing} for three-component composites case.
However, eigenvalues $s_n$ still lie in $[0,1]$ for $0 \le \eta({\bf r})
\le 1$.

We are now in the position to find an analytical representation for the
effective dielectric constant $\epsilon_{e}$ according to Eq.~(\ref{eq:1}). We
take advantage of Green's theorem, the boundary condition $\phi_{0}=-z$,
and the Maxwell equation ${\bf \nabla}\cdot {\bf D}=0$ to obtain the
effective dielectric constant
\begin{eqnarray}\label{tfeps_e}
\frac{\epsilon_e}{\epsilon_2} 
 &=& \frac {1}{\epsilon_2 V}\int \left(-\nabla\phi\right)\cdot {\bf D} dV\nonumber\\
 &=&\frac {-1}{V}\int \hat{\bf z} \cdot \left[ \left(
     1-\frac{1}{s} \eta({\bf r})\right) \nabla\phi
      \right] dV \\
 &=& 1+ \frac{1}{sV} \langle z | \phi\rangle\nonumber.
\end{eqnarray}
If we now introduce the reduced response~\cite{Bergman}
\begin{equation}
F(s)=1-\frac{\epsilon_e}{\epsilon_2},
\label{eq:13}
\end{equation}
and substitute Eq.~(\ref{solution}) into Eq.~(\ref{tfeps_e}) we find
\[
F(s)
= \frac{1}{V} \sum_{n}
                \frac{\left| \langle z|\phi_n\rangle \right| ^2}
                     { \langle \phi_n|\phi_n\rangle}
                \left(\frac {1}{s-s_n}\right).
\]
We can now express the effective dielectric constant as 
\begin{equation}\label{effective}
\epsilon_e=\epsilon_2 \left (1- \sum_{n}\frac{f_n}{s-s_n}\right),
\end{equation}
where $f_n$ is given by
\[
f_n=\frac{1}{V}\frac{\left| \langle z|\phi_n\rangle \right| ^2}
{ \langle \phi_n|\phi_n\rangle}.
\]
Using the above equations, we obtain the following sum rule 
\begin{eqnarray}
\sum_{n}f_{n}&=&\frac{1}{V}\langle z|z \rangle\nonumber\\
             &=&\frac{1}{V}\int d V \eta({\bf r}) \nabla z \cdot \nabla z\nonumber\\
             &=&\frac{1}{V}\int d V \eta({\bf r}).
\label{eq:17}
\end{eqnarray}

It is worth noting that the sum rule will not equal to the volume
fraction of inclusion. This is different from the Bergman-Milton spectral
representation for two homogeneous systems, in which the sum rule equals
to the volume fraction of the inclusion.

\begin{figure}[tbh]
\centerline{\epsfig{file=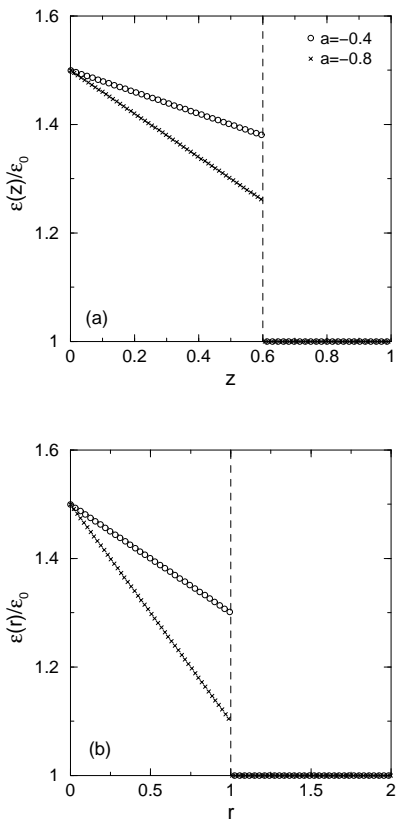,width=150pt}}
\caption{(a) Dielectric profile of  various graded films at $h=0.6$. 
(b) Dielectric proifle of various graded sphere with unit radius. 
Parameters used: $\epsilon_{2}=1$ and $s=-2$.}
\label{fig:1}
\end{figure}

When the operator $\hat{\Gamma}$ has a continuous spectrum,
Eq.~(\ref{effective}) should be replaced with the integral form
\begin{equation}
\epsilon_e=\epsilon_2 \left(1- \int ds^\prime \frac{m(s^\prime)}{s-s^\prime} \right),
\label{eq:18}
\end{equation}
where $m(s^\prime)$ is the spectral density function. Then, the reduced
response becomes
\begin{equation}
F(s)=\int ds^{\prime}\frac{m(s^\prime)}{s-s^\prime}.
\label{eq:19}
\end{equation}
If we write $s$ as $s+i0^{+}$, the right side of Eq.~(\ref{eq:19}) becomes 
\[
P\int ds^{\prime}\frac{m(s^\prime)}{s-s^\prime}-i\pi m(s),\nonumber
\]
and thus, $m(s^\prime)$ is given through the limiting process
\begin{equation}
m(s^\prime)=-\frac{1}{\pi}\mathrm{Im}[F(s^\prime+i0^{+})].
\label{eq:21}
\end{equation}

This final result is identical in form to Bergman's expression for the
analogous function in scalar composite materials.  However, there are
differences in the derivation, namely, the definition of the inner
product Eq.~(\ref{scalPr}), the continuous graded microstructure
$\eta({\bf r})$, the sum rule, as well as the range of eigenvalues $s_n$.

From Eq.~(\ref{eq:18}) it is evident that if the spectral density function
${\mathit m}({\mathit s}^{\prime})$ is known, the effective dielectric
constant can be obtained accurately, and vice versa. The spectral
representation has been used to analyze the effective dielectric
properties of composites. Recently, Levy and Bergman~\cite{Levy} also
used it in their study of nonlinear optical susceptibility. In this
regard, Sheng and coworkers~\cite{Ma} developed a practical algorithm for
calculating the effective dielectric constants based on the spectral
representation. In what follows, we restrict ourselves to a graded
composite both in one dimension and three dimensions, as well as
corresponding multilayer composites.

\section{Spectral density function of graded composites \label{sec:spectral}}
\subsection{Spectral density function of a graded film}

\begin{figure}[tbh]
\centerline{\epsfig{file=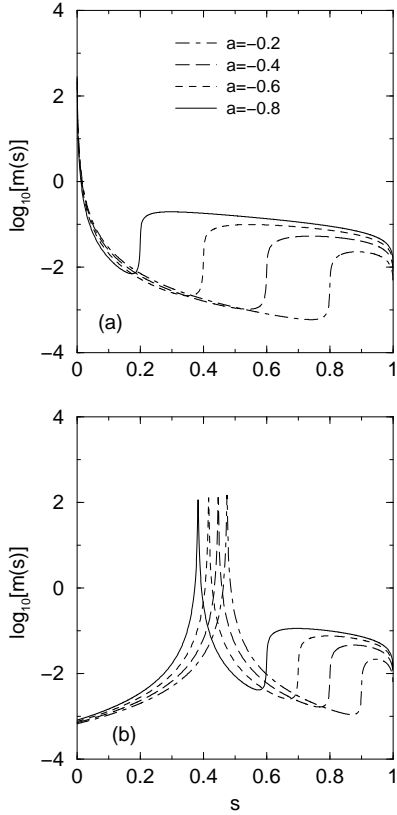,width=150pt}}
\caption{(a) Spectral density function of a graded film without an interface, 
i.\,e., ${\mathrm h}=1.0$. (b) Spectral density function of a graded 
film meeting a homogeneous medium at an interface ${\mathit h}=0.5$, and $\epsilon_{2}=1$.}
\label{fig:2}
\end{figure}

We consider a graded dielectric film of width ${\mathit L}$ , in which
two media meet at a planar interface as shown in Fig.~\ref{fig:1}\,(a). The first
medium $\epsilon_{1}(z)$ varies along $z-$axis, while the second medium
$\epsilon_2$ is homogeneous. We define the graded microstructure as
\begin{equation}
\eta(z)=\left\{\begin{array}{cc}
1+{\mathit a}z  \ \ &0<z \le {\mathit h}, \\
0 \ \  & {\mathit h}< z<{\mathit L},
\end{array} 
\right.
\label{eq:22}
\end{equation}
where ${\mathit a}$ and ${\mathit h}$ are real constants. They can be
varied to describe different graded films.  Thus, according to Eq.~(\ref{eq:3}),
the dielectric function of graded film can be expressed as
\begin{equation}
\epsilon(z)=\epsilon_{2}\left(1-\displaystyle 
\frac{\eta(z)}{{\mathit s}}\right).
\label{eq:23}
\end{equation}

Owing to the simple geometry of a graded film, we can use the equivalent
capacitance of a series combination to calculate the effective dielectric
constant as
\begin{equation}
\frac{1}{\epsilon_{e}}=\frac{1}{{\mathit L}}\int_{0}^{{\mathit L}}\frac{1}{\epsilon(z)} {\mathit d}z.
\label{eq:24}
\end{equation}
Substituting Eqs.~(\ref{eq:22}) and (\ref{eq:23}) into Eq.~(\ref{eq:24}), we obtain
\[
\frac{1}{\epsilon_e} =\displaystyle \frac{1 - {\mathit h}}{\epsilon_2} 
+ \displaystyle \frac{s \left[\ln
\left (1 - \displaystyle \frac{\eta(0)}{s}\right )
-\ln
\left(1 - \displaystyle \frac{\eta(h)}{s}\right)\right]}{{\mathit a}\epsilon_{2}},
\]
with the assumption  $L=1$.

We are now in a position to extend the Bergman-Milton spectral
representation of the effective dielectric constant~\cite{Bergman,Milton}
to a graded film. For a graded system, $\eta(z)$ can be a continuous
function in the inclusion medium. Using Eqs.~(\ref{eq:18})$-$(\ref{eq:21}), we obtain the
spectral density function for a graded film as
\begin{widetext}
\[
{\mathit m}({\mathit s}^{\prime})=-\displaystyle 
\frac{{\mathit a}s^{\prime} \mathrm{arg}
\left(\displaystyle \frac{s-1}{s-ah-1}\right)}
{\pi\left[\left( s^{\prime}\mathrm{arg}
\left(\displaystyle \frac{s-1}{s-ah-1} \right)\right)^{2}
+\left({\mathit a}({\mathit h}-1)
-s^{\prime} \ln \left(\displaystyle 
\frac{s^{\prime}-1}{s\prime -ah-1}\right) \right)^{2}\right]},
\]
\end{widetext}
where ${\mathit s}^{\prime}=\mathrm{Re}[{\mathit s}]$ and 
$\mathrm{arg}[\cdots]$ denote the arguments of complex functions.  

\subsection{Spectral density function of a graded sphere}

The above theory can be generalized to graded composites in three
dimensions. We consider a graded sphere with dielectric constant
$\epsilon_{1}(r)$ embedded into a homogeneous host medium with dielectric
constant $\epsilon_2$. The dielectric constant of the graded sphere
$\epsilon_1(r)$ varies along the radius $r$. We can obtain the effective
dielectric constant of a graded sphere using the spectral representation.  
We consider the graded microstructure as
%
\[
\eta(r)=\left\{\begin{array}{cc}
                  1+{\mathit a}r \ \ & 0<r \le {\mathit R},\\
                  0              \ \  & r>{\mathit R},
                 \end{array}\right. 
\]
where ${\mathit R}$ is the radius of the graded sphere. Thus, from
Eq.~(\ref{eq:3}) the dielectric constant in the graded sphere is given by
\begin{equation}
\epsilon(r)=\epsilon_{2}\left( 1-\frac{\eta(r)}{s}\right).
\label{eq:28}
\end{equation}
In the dilute limit the effective dielectric constant of a small volume
fraction ${\mathit p}$ of graded spheres embedded in a host medium is
given by~\cite{YuHui,Zhang}
\begin{equation}
\epsilon_{e}=\epsilon_{2}+3 \epsilon_{2} {\mathit p}{\mathit b},
\label{eq:29}
\end{equation}
where ${\mathit b}$ is the dipole factor of graded spheres embedded in a
host as given in Ref.~\cite{Dong}. Using Eq.~(\ref{eq:13}) and Eq.~(\ref{eq:29}), the
reduced response can be obtained as
\begin{equation}
{\mathit F}({\mathit s})=-3 \epsilon_{2} {\mathit p}{\mathit b}.
\label{eq:30}
\end{equation}
Thus, the spectral density function of a graded sphere can be given
through a numerical evaluation of Eq.~(\ref{eq:21}).

\section{spectral density function of multilayer composites \label{sec:multi}}

A multilayer composite is a special case of graded composites. The
gradation becomes continuous as the number of layers approaches infinity.  
To investigate the multilayer effect, we shall use a finite difference
approximation for the graded profile (Eqs.~(\ref{eq:23}) and (\ref{eq:28})) for a finite
number of layers. To mimic a multilayer system, we divide the interval
$[0,1]$ into $N$ equally spaced sub-intervals, $[0,z_1]$, $(z_1,z_2]$,
$\cdots$, $(z_N-1,1]$. Then we adopt the midpoint value of $\epsilon(z)$
for each sub-interval as the dielectric constant of that sublayer. In
this way, we calculate the effective dielectric constant, eigenvalues, as
well spectral density function for each $N$. It is worth noting that the
results of $N\to \infty$ (e.\,g., $N=1024$ recovers the results of graded
composites.

In addition to multilayer films, we can use the above approach to study
the much simpler problem of a two-layer film. In this system, we have two
layers of dielectric constants $\epsilon_1$, $\epsilon_2$, and host
$\epsilon_0$. Thickness are $hy$, $h(1-y)$, and $1-h$, respectively,
where $y$ is the length ratio between component $\epsilon_1$ and
component $\epsilon_{2}$.  We also define two microstructure parameters,
$\eta_1$ and $\eta_2$. If we let $s=1/(1-\epsilon_1/\epsilon_0)$, then
$\eta_1=1$, and
$\eta_2=(\epsilon_{0}-\epsilon_{2})/(\epsilon_{0}-\epsilon_{1})$.
According to Eq.~(\ref{eq:23}), the effective dielectric constant of the two-layer
film is now given by
\[
\frac{1}{\epsilon_e}=\frac{hy}{\epsilon_1}+\frac{h(1-y)}{\epsilon_2}+\frac{1-h}{\epsilon_0}.
\]
According to Eq.~(\ref{eq:13}), the reduced response can be given by
\begin{equation}
F(s)=\frac{F_1}{s-s_1}+\frac{F_2}{s-s_2},
\label{eq:32}
\end{equation}
where 
\begin{eqnarray}
F_1&=&\frac{h(s_1(y-y\eta+\eta)-\eta)}{s_1-s_2},\nonumber\\
F_2&=&-\frac{h(s_2(y-y\eta+\eta)-\eta)}{s_1-s_2},\nonumber\\
s_1&=&\frac{1}{2}\left[ 1-h(y-y\eta+\eta)+\eta \right. \nonumber\\
   & &  \left. -\sqrt{4\eta (-1+h)+(1-h(y-y\eta+\eta)+\eta)^2} \right],\nonumber\\
s_2&=&\frac{1}{2}\left[ 1-h(y-y\eta+\eta)+\eta \right. \nonumber\\
   & &  \left.     +\sqrt{4\eta (-1+h)+(1-h(y-y\eta+\eta)+\eta)^2} \right].\nonumber
\end{eqnarray}
From the sums of $F_1$ and $F_2$ and the integral of graded
microstructure $\eta(z)$ given by Eq.~(\ref{eq:22}), we can check that the sum
rule expressed by Eq.~(\ref{eq:17}) is obeyed. It should also be noted that there
are two poles in the expression for the reduced response corresponding to
two peaks in the spectral density function. If $h=1$, then $s_{1}=0$,
that is, one peak is located at zero, which is explicitly shown in
Fig.~\ref{fig:2}(a).

\begin{figure}[tbh]
\centerline{\epsfig{file=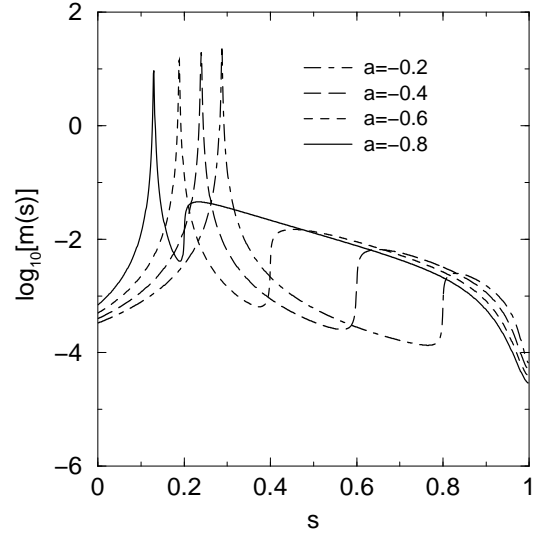,width=200pt}}
\caption{(Spectral density function of a graded sphere with volume fraction ${\mathit p}=0.1$.}
\label{fig:3}
\end{figure}

\begin{figure*}[tbh]
\centerline{\epsfig{file=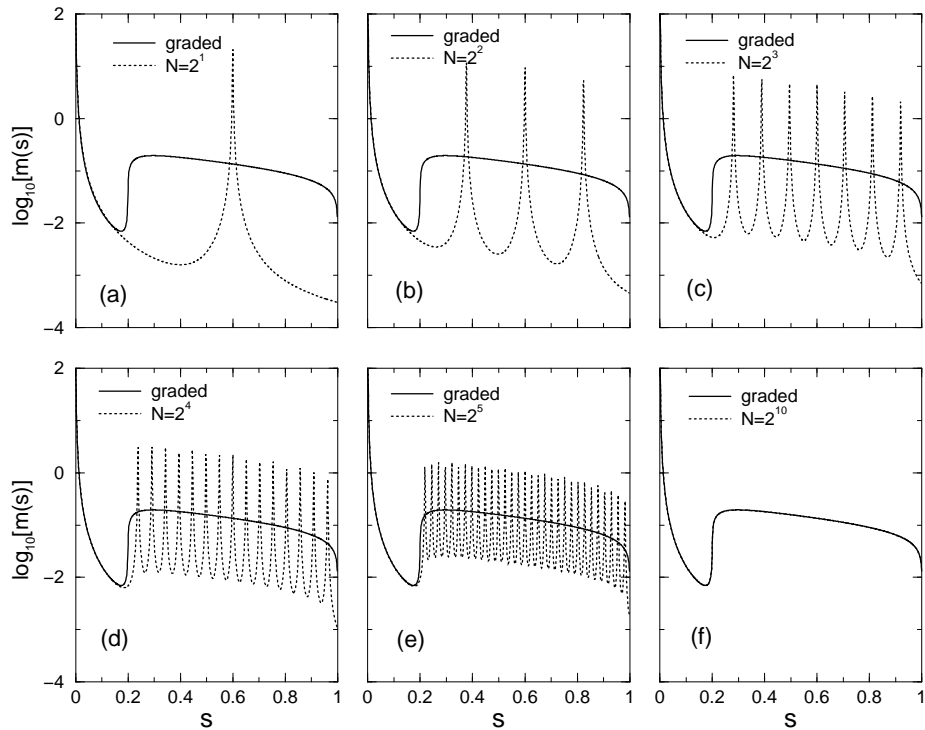,width=350pt}}
\caption{Spectral density function of various multilayer film with $\epsilon_0=1$ and $a=-0.8$.}
\label{fig:4}
\end{figure*}

Similarly, we can also apply our graded spectral representation to a
single-shell sphere of core dielectric constant $\epsilon_1$, covered by
a shell of $\epsilon_2$, and suspended in a host of $\epsilon_0$. In this
example, we can also define two microstructure parameters $\eta_1$ and
$\eta_2$. If we let $s=1/(1-\epsilon_1/\epsilon_0)$, then $\eta_1=1$, and
$\eta_2=(\epsilon_{0}-\epsilon_{2})/(\epsilon_{0}-\epsilon_{1})$. The
dipole factor of single-shell sphere is given~\cite{YuHui,Zhang}
%
\[
b=\frac{\epsilon_{2}-\epsilon_{0}+
(\epsilon_{0}+2\epsilon_{2})xf^3}{\epsilon_{2}+2\epsilon_{0}+2
(\epsilon_{2}-\epsilon_{0})xf^3},
\]
%
where $f$ is the ratio between radius core and radius shell, and $x$ is
given by
%
\[
x=\frac{\epsilon_{1}-\epsilon_{2}}{\epsilon_{1}+2\epsilon_{2}}.
\]
%
Then, we can also write Eq.~(\ref{eq:30}) similarly to Eq.~(\ref{eq:32}), where the
residues and eigenvalues are given by,
\begin{eqnarray}
F_1&=&\displaystyle \frac{-3 p s_1 [(-1+\eta)y^3-\eta]-\eta p [1-2(-1+\eta)y^3+2 \eta]}{3(s_1-s_2)},\nonumber\\
F_2&=&\displaystyle \frac{3 p s_2 [(-1+\eta)y^3-\eta]+\eta p [1-2(-1+\eta)y^3+2 \eta]}{3(s_1-s_2)},\nonumber\\
s_1&=&\frac{1}{6}\left[ 1+3\eta-\sqrt{1+(2-8y^3)\eta+(1+8y^3)\eta^2}\right],\nonumber\\
s_2&=&\frac{1}{6}\left[ 1+3\eta+\sqrt{1+(2-8y^3)\eta+(1+8y^3)\eta^2}\right].\nonumber
\end{eqnarray}
Analysis shows that the spectral representation for $N=2$ contains two
simple poles corresponding to two peaks in the spectral density function.
Therefore, we draw the conclusion that, $N$ peaks are a result of $N$
layers. Moreover, $N-1$ peaks will accumulate into a continuous broad
absorption spectrum when $N$ tends toward infinity, which can be seen
from Fig.~\ref{fig:4}(f)~and~\ref{fig:5}(f).

\section{Numerical results  \label{sec:numerical}} 

We are now in a position to do some numerical calculations of the
spectral density function from Eqs.~(\ref{eq:18}) and (\ref{eq:21}). A small but finite
imaginary part in the complex parameter has been used in the
calculations. Without any loss of generality, we choose ${\mathit L}=1$
and ${\mathit R}=1$ for convenience. We show the effect of different
graded profiles, as well as the effect of the thickness of the inclusion.  
It should be noted, that in all figures the range of $s$ is limited to
$[0,1]$, because we chose $-1<a<0$ which limits the value of $\eta$ into
$[0,1]$.

Fig.~\ref{fig:1} displays the dielectric profile of a graded film (Fig.~\ref{fig:1}(a)) and a
graded sphere (Fig.~\ref{fig:1}(b)). This figure obviously shows that the dielectric
constant varies with the position in inclusion while a constant in host
medium. Also, different values of $a$ accord with different graded
materials.

In Fig.~\ref{fig:2}(a), we plot the spectral density function ${\mathit m}({\mathit
s})$ of a graded film without an interface against the spectral parameter
for various graded microstructures $\eta(z)$. It is evident that there is
always a broad continuous band in the spectral density function. Both the
strength as well as the width of the continuous part of ${\mathit
m}({\mathit s})$ increase with the gradient of the dielectric profile.
Thus, the previous results of the broad surface-plasmon band can be
expected. Note that there is a sharp peak at ${\mathit s}=0$, which is
also present in a homogeneous film. In Fig.~\ref{fig:2}(b), we plot the spectral
density function of a graded film meeting a homogeneous medium at an
interface for various graded microstructure $\eta(z)$. Again, there is
always a broad continuous band in the spectral density function. However,
the sharp peak has now shifted to a finite value of ${\mathit s}$, which
is also present in a homogeneous film.

In Fig.~\ref{fig:3}, the spectral density function of graded sphere is displayed
for a volume fraction ${\mathit p}=0.1$. In this case, the interface
always exists. It is clear that a broad continuous function in the
spectral density function is always observed, as well as the shift of the
sharp peak. However, the decrease of the broad continuous function is
more abrupt for graded sphere than for graded film with increasing
${\mathit s}$.

\begin{figure*}[tbh]
\centerline{\epsfig{file=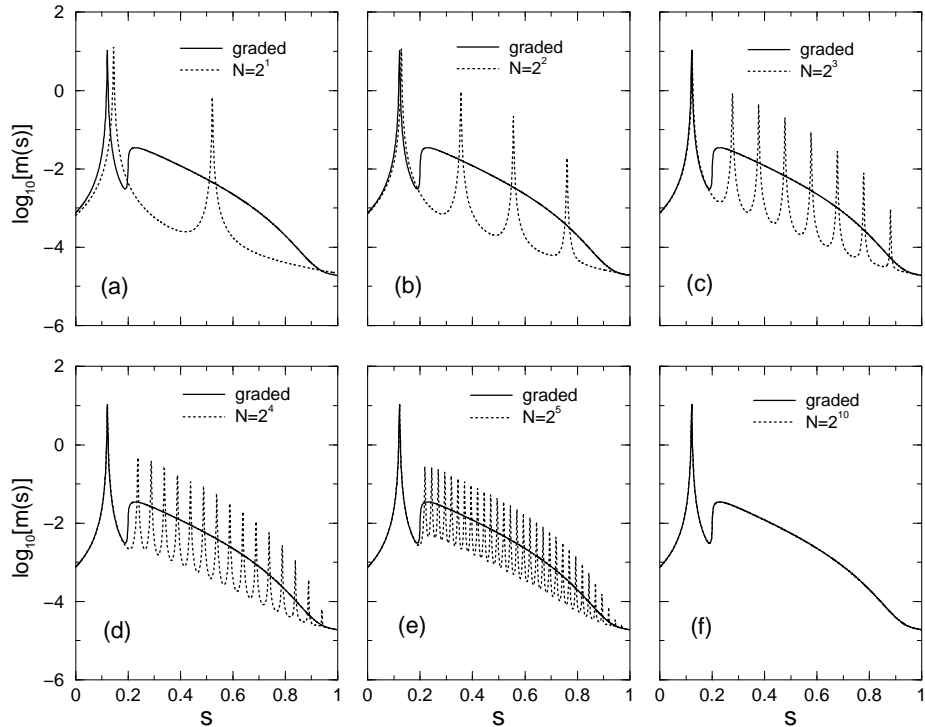,width=350pt}}
\caption{Spectral density function of various multi-shell sphere with $\epsilon_0=1$, $a=-0.8$ and $p=0.1$.}
\label{fig:5}
\end{figure*}

Figures~\ref{fig:4}~and~\ref{fig:5} display the spectral density function for a multilayer
film and a sphere, respectively. It is clear that there are always $N$
sharp peaks for $N$ layers. Moreover, it is worth noting that there
occurs a transition from sharp peaks to a broad continuous band with
increasing $N$ (see Fig.~\ref{fig:4}(f) and Fig.~\ref{fig:5}(f)), that is, the graded results
are recovered by the limit results of $N\to \infty$. In particular, we
had obtained the analytical expression of spectral density function for
$N=2$. There are two resonances corresponding to the two peaks in
Fig.~\ref{fig:4}(a) and Fig.~\ref{fig:5}(a).

\section{Discussion and Conclusion  \label{sec:discussion}}

We have investigated a graded composite film and a sphere by means of the
Bergman-Milton spectral representation. It has been shown that the
spectral density function can be obtained analytically for a graded
system. However, unlike in the case of homogeneous constituent
components, the characteristic function is a continuous function due to
the presence of gradation. Moreover, the derivation as well as some
salient properties, namely, the sum rule, the definition of inner
product, the definition of the integral-differential operators, and the
range of spectral parameters, do change because of the continuous
variation of the dielectric profile within the constituent components.  
It should be noted that in graded composite, the eigenvalues are not
limited to $[0,1]$, and they can be extended to $-\infty \le s_n \le
\infty$ for the full region $\eta$, i.\,e., $-\infty \le \eta \le
\infty$. In this work, however for simplicity, we investigated the
spectral density function in $0 \le s \le 1$ by choosing $-1 < a <0$ to
limit the value of $\eta$ into $[0,1]$.

We also study multilayer composites and calculated the spectral density
function versus the number of layers, to explicitly demonstrate that the
broad continuous spectrum arises from the accumulation of poles when the
number of layers tends to infinity. This finding coincides with the broad
surface-plasmon absorption band associated with the optical properties of
graded composites.

To sum up, we have investigated the spectral density function of graded
film and graded sphere, as well as multilayer cases. There is always a
broad continuous function in the spectral density function in graded
composite, but simple poles in multilayer composite, the number of pole
depends on the number of layers. Moreover, there is a gradual transition
from sharp peaks to a broad continuous band until the graded composite
results recover in the limit of $N\to \infty$.

\begin{acknowledgments}

This work was supported by the Research Grant Council of Hong Kong SAR
Government Earmarked Grant (K.\,W.\,Y.) 
and by the Academy of Finland (L.\,D., M.\,K.).

\end{acknowledgments}

\end{document}